\documentclass[a4paper]{jpconf}

\usepackage[dvips]{color}
\usepackage{graphicx}
\usepackage{amsmath,amssymb}
\usepackage{epsfig}

\begin{document}

\newcommand{\bnabla}{\mbox{\boldmath $\nabla$}}
\newcommand{\brho}{\mbox{\boldmath $\rho$}}
\newcommand{\bea}{\begin{eqnarray}}
\newcommand{\eea}{\end{eqnarray}}
\newcommand{\eg}{{e.g., }}
\newcommand{\ie}{{i.e., }}
\newcommand{\spaceint}[2]{\int_{#1} d^3 #2 \;}
\newcommand{\vect}[1]{\mathbf{#1}}
\newcommand{\vat}{V^{\rm att}}
\newcommand{\di}{\displaystyle}
\newcommand{\rp}{r_{||}}
\newcommand{\ep}{\varepsilon_\Pi}
\newcommand{\ef}{\varepsilon_F}
\newcommand{\emix}{\varepsilon_{\rm m}}
\newcommand{\ev}{\varepsilon_{\rm rep}}
\newcommand{\Ezp}{E_{z,0^+}}
\newcommand{\Ezm}{E_{z,0^-}}
\newcommand{\Ep}{E_{||,0}}
\newcommand{\ro}{r_{0,{\rm ref}}}

\title{Sound-mediated dynamic correlations between colloidal particles
  in a quasi-one-dimensional channel}

\author{D Frydel and H Diamant}

\address{Raymond and Beverly Sackler School of Chemistry, Tel Aviv University, 
Tel Aviv 69978, Israel}

\ead{hdiamant@tau.ac.il}

\begin{abstract}
  We study the hydrodynamic interactions between colloids suspended in
  a compressible fluid inside a rigid channel. Using
  lattice--Boltzmann simulations and a simplified hydrodynamic theory,
  we find that the diffusive dynamics of density perturbations (sound)
  in the confined fluid give rise to particle correlations of
  exceptionally long spatial range and algebraic temporal decay. We
  examine the effect of these sound-mediated correlations on
  two-particle dynamics and on the collective dynamics of a
  quasi-one-dimensional suspension.
\end{abstract}

\section{Introduction}
%=====================
\label{sec_intro}

Particles embedded in an unbounded fluid are dynamically correlated by
hydrodynamic interactions that decay like $1/r$ (where $r$ denotes the
separation between particles) \cite{HappelBrenner}. This result is
understood in terms of the fundamental solution to the steady-state
Stokes equation, $\eta\nabla^2{\bf u}=\bnabla p-{\bf b}\delta({\bf
  r})$, supplemented by the fluid-incompressibility condition
$\bnabla\cdot{\bf u}=0$. Here ${\bf u}$ is the flow velocity field,
$p$ the pressure field, ${\bf b}$ a point force, and $\eta$ the shear
viscosity. It follows from the first equation that the vorticity,
$\bnabla\times{\bf u}$, satisfies the Laplace equation, leading in the
case of an unbounded fluid to the $1/r$ slow decay of the velocity
field. When particles are confined to a quasi-one-dimensional (q1D)
channel, however, these long-range dynamic correlations are cut off
\cite{Haim02}, the screening being attributed to dissipation at the
fluid-solid interface, \ie to the loss of fluid momentum at the
channel boundaries. The resulting screening length is comparable to
the channel width.

The dissipative effect of the boundaries can be incorporated through
an additional friction term in an ``effective fluid'' model
\cite{Brinkman,Mazenko82}, $\eta\nabla^2{\bf u}=\rho_0\xi{\bf
  u}+\bnabla p-{\bf b}\delta({\bf r})$. In the extra term $\rho_0$ is
the fluid mass density and $\xi$ an effective friction coefficient.
The parameter $\xi$ has units of inverse time and characterizes the
rate of momentum loss to the boundaries. It can be estimated,
therefore, as the inverse of the time it takes fluid momentum to
diffuse to the channel boundary, $\xi\sim\nu/h^2$, $\nu=\eta/\rho_0$
being the kinematic shear viscosity and $h$ the channel width. This
simplified, phenomenological approach has been shown by simulations
\cite{Frenkel97,Frenkel98,Frenkel99,Frydel07} and analytical
calculations \cite{FelderhofJCP11} to correctly reproduce the
qualitative behavior of confined fluids.  The modified Stokes equation
leads to a Helmholtz equation for the vorticity, whose regular
solutions are exponentially screened functions. Yet, the friction that
suppresses the vorticity (transverse fluid stress) does not similarly
suppress the pressure field (longitudinal fluid stress) emanating from
a locally applied force.  Consequently, the application of a localized
force generates a long-range pressure distribution, which may give
rise to long-range flows and forces on embedded particles. For
example, in a q2D suspension confined between two plates, the pressure
distribution creates hydrodynamic interactions that decay like
$1/r^2$, while the $1/r$ correlations due to transverse momentum
transfer are cut off \cite{Haim04}. Thus, although the hydrodynamic
interactions in the q2D system remain long-ranged, the very mechanism
for correlations changes \cite{jpsj09}. In the case of a q1D channel,
the pressure generated by a localized force far along the channel is
constant and, therefore, does not produce long-range forces.  Thus,
within the Stokes description of a steady incompressible flow, in a
q1D rigid channel long-range hydrodynamic interactions arise neither
from transverse stresses nor from longitudinal ones, and the
correlation between particles is screened, the screening length being
proportional to the channel width \cite{Haim02}.

The omission of compressibility, and with it of compression
(longitudinal sound) modes, is usually justified for unbounded fluids,
where the effects of sound are short-lived. The short propagation time
to any reasonable distance $r$, $\tau_s\sim r/c_s$, is a result of the
large speed of sound, $c_s \sim 10^3$ m/s for ordinary liquids such as
water. The effect of sound on the short-time hydrodynamic interactions
in unconfined suspensions was addressed in earlier works
\cite{Ladd1995,Bakker2002,Henderson2002}. The corresponding effects
are subtle and short-lived, quickly taken over by the fully
established hydrodynamic interaction due to transverse flow.

In confined geometries, however, the behavior of sound modes
qualitatively changes. Inclusion of fluid compressibility has been
found to lead to a velocity autocorrelation function of a confined
particle, which decays with time only algebraically, with a negative
long-time tail \cite{Frenkel97,Frenkel98,Frenkel99,FelderhofJFM10}.
[In the case of a q1D channel the decay is $\sim(-t^{-3/2})$.]  Thus,
under confinement, sound acquires long-time memory without a
well-defined relaxation time, and the incompressibility assumption
loses, at least in principle, its justification. Informed by this
result, one suspects that the long memory of sound modes may play a
role in the dynamic correlations between different particles in a
channel: since sound may propagate to long distances, the resulting
correlations will be long-ranged, and the algebraic temporal decay
will endow these interactions with a long-time memory. This novel
mechanism for correlations challenges the conventional wisdom
described above, according to which fluid-mediated correlations in a
rigid channel are exponentially screened.  The question that comes to
mind is whether the long-time memory of sound is sufficiently
significant to be observed experimentally and to undermine predictions
of the theories based on the assumption of incompressibility. We
remind that the algebraic temporal decay of sound modes in a channel
does not imply that sound is capable of causing steady rearrangement
of the fluid.  The integrated effect of the $t^{-3/2}$ tail decays as
$t^{-1/2}$, which is very slow, but still implies that at infinite
time (the steady-state limit) all perturbations due to sound
eventually vanish.

In a recent Letter \cite{Frydel10} we have demonstrated that the
heuristic arguments given above are correct\,---\,namely, the
diffusive, slowed-down sound modes of a fluid confined in a rigid
channel mediate long-range and long-time velocity correlations between
suspended particles. In the present publication we report detailed
findings concerning the compressive response of the fluid and its
effect on particles, which were not included in that brief
publication.  Two tools are employed and compared: simplified
analytical calculations and lattice--Boltzmann simulations. We assume
no-slip boundary conditions at the fluid-particle interface and
consider either slip or no-slip boundary conditions at the channel
boundaries. Fluid dynamics in a q1D channel strongly depends on the
boundary conditions imposed at the edges of the channel
\cite{Bhattacharya10}. In the calculations we assume an infinitely
long channel with a vanishing flow velocity at the edges. In the
simulation we take a sufficiently long channel, such that all the
temporal results presented below are insensitive to the system size.
Further details of the simulations can be found in
Ref.~\cite{Frydel10}.

\section{Dynamics of fluid density perturbations}
%================================================
\label{sec_sound}

\subsection{Unbounded fluid}
%---------------------------

We begin by revisiting the compressive response of an unbounded
Newtonian fluid to an impulsive force \cite{FelderhofPhysica10}, to
which we will later compare the case of a fluid confined in a channel
\cite{FelderhofJFM10,FelderhofJFM11}.

Isothermal fluid dynamics \cite{LL} is governed by the
momentum-conservation Navier--Stokes equation,
\begin{equation}
\rho\Big[\partial_t{\bf u}+({\bf u}\cdot\bnabla){\bf u}\Big]
=-{\bf\nabla}p+\eta\nabla^2{\bf u}
+(\eta_v+\eta/3)\bnabla(\bnabla\cdot{\bf u})+{\bf f},
\end{equation}
and the mass-conservation equation,
\begin{equation}
\partial_t\rho + \bnabla\cdot(\rho{\bf u}) = g.
\end{equation}
In these equations $\rho$ is the fluid density field, ${\bf f}$ an
external force density, and $\eta$ and $\eta_v$ are the shear and
volume viscosities, respectively. The function $g$ could represent,
for example, a mass monopole, $m\delta({\bf r}-{\bf r}_0)$, where at
the point ${\bf r}_0$ fluid is generated or lost, or a mass dipole,
$-{\bf d}\cdot\bnabla\delta({\bf r}-{\bf r}_0)$, where fluid is at the
same time created and lost, thus globally conserving mass but
generating a flow.

The hydrodynamic equations are linearized by neglecting the convective
term, $({\bf u}\cdot\bnabla){\bf u}$, and separating the thermodynamic
variables into equilibrium and small perturbation parts, $\rho({\bf
  r},t)=\rho_0+\delta\rho({\bf r},t)$ and $p({\bf r},t)\simeq p_0 +
(\partial p/\partial\rho)\delta\rho = p_0 + c_s^2\delta\rho({\bf r},t)$. The
resulting linearized equations read
\begin{equation}
\rho_0\partial_t{\bf u}
=-c_s^2{\bf\nabla}\delta\rho+\eta\nabla^2{\bf u}
+(\eta_v+\eta/3)\bnabla(\bnabla\cdot{\bf u})+{\bf f},
\label{eq:NS_lin}
\end{equation}
\begin{equation}
\partial_t\delta\rho + \rho_0\bnabla\cdot{\bf u} = g.
\end{equation}

When dealing with sound, it is convenient to decompose the linearized
equations into transverse and longitudinal components (the so called
Helmholtz decomposition), ${\bf u}={\bf u}^T+{\bf u}^L$, which satisfy
$\bnabla\cdot{\bf u}^T=0$ and $\bnabla\times{\bf u}^L=0$.  The
decomposition yields
\begin{equation}
\rho_0\partial_t{\bf u}^T=\eta\nabla^2{\bf u}^T+{\bf f}^T,
\label{eq:uT}
\end{equation}
%\begin{equation}
%\bnabla\cdot{\bf u}^T=0
%\label{eq:mT}
%\end{equation}
and
\begin{subequations}
\begin{align}
\label{eq:uL}
\rho_0\partial_t{\bf u}^L
=-c_s^2{\bnabla}\delta\rho+(\eta_v+4\eta/3)
\nabla^2{\bf u}^L+{\bf f}^L,\\
\label{eq:mL}
\partial_t\delta\rho + \rho_0\bnabla\cdot{\bf u}^L = g,
\end{align}
\label{eq:L}
\end{subequations}
where we have used the identity ${\bnabla}({\bnabla}\cdot{\bf
  u})=\nabla^2{\bf u}^L$.  After separating the force-density field
into transverse and longitudinal components, ${\bf f}={\bf f}^T+{\bf
  f}^L$, the two flows become independent. The sound is associated
with the longitudinal flow, which is coupled dynamically to the
density\,---\,\ie the sound modes characterize the propagation and
relaxation of a density perturbation.
%Subsequently, we are interested in the dynamics of longitudinal flow. 

The time evolution of a density perturbation is obtained by taking the
divergence of Eq.~(\ref{eq:uL}) and then transforming it using the
mass-conservation equation, Eq.~(\ref{eq:mL}),
\begin{equation}
\partial_t^2\delta\rho = c_s^2\nabla^2\delta\rho
+2\Gamma\nabla^2\partial_t\delta\rho
-{\bf b}\cdot{\bnabla}\delta({\bf r})\delta(t).
%-{\bf\nabla}\cdot{\bf f}^L
\label{eq:rho}
\end{equation}
In Eq.~(\ref{eq:rho})
$\Gamma=\frac{1}{2}\frac{\eta_v+4/3\eta}{\rho_0}$, and we have set
$g=0$ and the force density to an impulsive point force, ${\bf f}={\bf
  b}\delta({\bf r})\delta(t)$. Note that the application of the
impulsive point force generates a momentary mass dipole in the
equation for the density, $-({\bf b}\delta(t))\cdot{\bnabla}\delta({\bf r})$.
Equation (\ref{eq:rho}) represents a damped wave equation with a
damping parameter $\Gamma$.  If $\Gamma=0$, we retrieve a wave
equation for $\delta\rho$, and if $c_s=0$, we obtain a diffusion
equation for $\partial_t\delta\rho$.  Upon Fourier-transforming the
spatial coordinates, Eq.~(\ref{eq:rho}) becomes
$$
\partial_t^2\delta\rho_k + (2\Gamma k^2)\partial_t\delta\rho_k 
+ (c_s^2k^2)\delta\rho_k = -i{\bf k}\cdot{\bf b}\delta(t),
$$
whose solution reads
\begin{equation}
\delta\rho_k(t) = \frac{-i{\bf k}\cdot{\bf b}
e^{-\Gamma k^2t}}{\sqrt{c_s^2k^2-\Gamma^2k^4}}
\sin\Big(t\sqrt{c_s^2k^2-\Gamma^2k^4}\Big).
\label{eq:rho_k}
\end{equation}
If $c_s^2-\Gamma^2k^2 > 0$, the density perturbation propagates as a
decaying wave.  If, on the other hand, $c_s^2-\Gamma^2k^2 < 0$, the
sine becomes a hyperbolic sine, and the density perturbation
propagates diffusively.  The crossover between these two behaviors
occurs above a wavelength $\sim \Gamma/c_s$.  For water
$\Gamma/c_s\approx 1$~nm, while a typical colloid diameter is
$\sigma\approx 10^3$~nm. Therefore, particle correlations arising from
sound are mediated by essentially underdamped wave-like modes. In this
large-wavelength limit, $k^{-1}\gg\Gamma/c_s$, Eq.~(\ref{eq:rho_k})
further simplifies to
\begin{equation}
\delta\rho_k(t)\approx \frac{-i{\bf k}\cdot{\bf b}
e^{-\Gamma k^2t}}{c_sk}\sin(c_skt).
\label{eq:rho_k2}
\end{equation}

\begin{figure}[tbh]
 \center{\epsfig{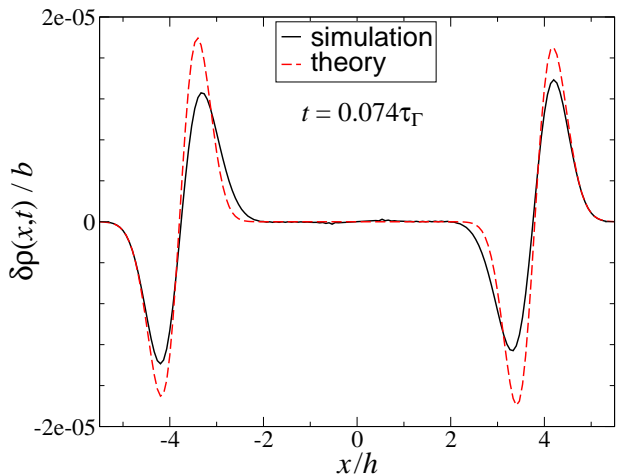}}
 \caption{Snapshot of the density perturbation generated by an impulsive 
   force in an unbounded fluid. The force is applied at the origin in
   the positive $x$ direction. Lattice--Boltzmann simulation result
   (where the force is applied to a rigid sphere of diameter $\sigma$)
   is compared with the analytical one for a point force
   [Fourier-inverted Eq.~(\ref{eq:rho_k})]. The snapshot is taken
   $0.074\tau_\Gamma$ after the impulse, where
   $\tau_{\Gamma}=h^2/\Gamma$ and $h=1.5\sigma$.}
\label{fig:rho_b}
\end{figure}

Figure~\ref{fig:rho_b} shows a snapshot of the density perturbation,
generated a certain given time after the application of an impulsive
force inside an unbounded fluid. We compare the result of a
lattice--Boltzmann simulation, where the force is applied to a rigid
spherical particle, with Eq.~(\ref{eq:rho_k}) for a point force.  The
analytical calculation reproduces all the features of the simulation,
with a quantitative discrepancy that can be ascribed to the finite
particle size in the simulation.

\subsection{Channel with slip boundary conditions}
%-------------------------------------------------

\begin{figure}[tbh]
 \center{\epsfig{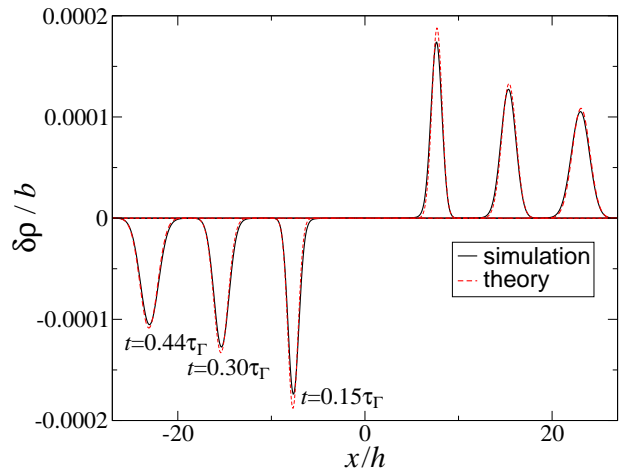}}
 \caption{Three snapshots of a density perturbation for a fluid in a square 
   channel with slip boundary conditions. An impulsive force is
   applied along the channel axis in the positive $x$ direction.
   Lattice--Boltzmann simulation results (where the force is applied
   to a rigid sphere of diameter $\sigma$) are compared with the
   analytical ones for a force distributed uniformly over the square
   cross-section of side $h=1.5\sigma$ [Eq.~(\ref{eq:rho1D})]. The
   time unit is $\tau_{\Gamma}=h^2/\Gamma$.}
 \label{fig:rhot_Q1D_impulse_slip}
\end{figure}

For a fluid confined in a channel with slip boundary conditions, we
neglect variations of the density over the channel cross-section and
consider only the direction along the channel, $x$.
Equation~(\ref{eq:rho}) is then rewritten as
$$
\partial_t^2\delta\rho=c_s^2\delta\rho''+2\Gamma\delta\partial_t\rho''
-\frac{b}{h^2}\delta'(x)\delta(t),
$$
where a prime indicates differentiation with respect to $x$, and an
impulse is evenly distributed over the cross--sectional area of a
square channel of side $h$. In the long wavelength limit the solution
takes the form,
\begin{equation}
\delta\rho\approx \frac{b}{2c_sh^2}\Bigg(
\frac{e^{-(x-c_s t)^2/(4\Gamma t)}}{\sqrt{4\pi\Gamma t}}-
\frac{e^{-(x+c_s t)^2/(4\Gamma t)}}{\sqrt{4\pi\Gamma t}}\Bigg).
\label{eq:rho1D}
\end{equation}
There are two signals propagating with the speed of sound in opposite
directions. In addition, each signal spreads out diffusively with
diffusion coefficient $\Gamma$. Figure \ref{fig:rhot_Q1D_impulse_slip}
shows density perturbation snapshots generated by an impulse applied
at the center and along a square channel with slip boundary
conditions. Lattice--Boltzmann simulation results, where the impulse
is applied to a spherical particle, agree well with
Eq.~(\ref{eq:rho1D}).

\subsection{Channel with no-slip boundary conditions}
%----------------------------------------------------

To account for no-slip boundary conditions at the channel walls, we
introduce an effective friction term, $(-\rho_0\xi{\bf u}^L)$, to the
right-hand side of Eq.~(\ref{eq:uL}). This leads to an extra term,
$(-\xi\partial_t\delta\rho)$, in Eq.~(\ref{eq:rho}). Thus, neglecting
again density variations transverse to the channel axis, we have the
effective 1D equation,
$$
\partial_t^2\delta\rho=c_s^2\delta\rho''+2\Gamma\partial_t\delta\rho''
-\xi\partial_t\delta\rho-\frac{b}{h^2}\delta'(x)\delta(t).
$$
For the friction coefficient (rate of momentum loss to the walls)
we take $\xi=\alpha\nu/h^2$, where $\nu=\eta/\rho_0$ is the kinematic
shear viscosity, and $\alpha$ a geometrical prefactor, dependent on
the shape of the channel cross-section. (The value of $\alpha$ can be
determined from the resistance of the channel to a steady
pressure-driven flow, \eg $\alpha\simeq 28.454$ for a square
cross-section and $\alpha=36$ for a circular one
\cite{HappelBrenner,FelderhofJCP11}.)

In Fourier space the solution reads
\begin{equation}
\delta\rho_k(t) = \frac{b}{h^2}\frac{(-ik)e^{-(\Gamma k^2+\alpha\nu/2h^2)t}}
{\sqrt{c_s^2k^2-(\Gamma k^2+\alpha\nu/2h^2)^2}}
\sin\Big(t\sqrt{c_s^2k^2-(\Gamma k^2+\alpha\nu/2h^2)^2}\Big).
\label{eq:rhoxi_k}
\end{equation}
The corresponding signal propagates with underdamped oscillations only
for a narrow range of small wavelengths, satisfying
\begin{equation}
\frac{c_s}{2\Gamma}\Bigg(1-\sqrt{1-\frac{2\Gamma}{D_s}}\Bigg)
<k<
\frac{c_s}{2\Gamma}\Bigg(1+\sqrt{1-\frac{2\Gamma}{D_s}}\Bigg),
\label{eq:limits}
\end{equation}
where $D_s=c_s^2h^2/(\alpha\nu)$ is the sound diffusion coefficient.
Hence, the friction at the walls transforms the large-scale dynamics
from oscillating to diffusive.
%For water in a microchannel $\alpha_c\approx 17860$ and for our 
%simulation $\alpha_c=47.5$.  
In the large-wavelength limit Eq.~(\ref{eq:rhoxi_k}) reduces to
\begin{equation}
\delta\rho_k(t) \approx\Big(\frac{b}{\alpha\nu}\Big){(-ik)}e^{-D_sk^2t},
\label{eq:rhoxi_smallk}
\end{equation}
which in real space reads,
\begin{equation}
\delta\rho\approx\frac{b}{\alpha\nu}\frac{2}{\sqrt{\pi}}
\frac{xe^{-x^2/4D_st}}{(4 D_s t)^{3/2}}.
\label{eq:rho1D_xi}
\end{equation}
Thus, at large distances the density perturbation evolves as a
diffusive dipolar signal.  

\begin{figure}[tbh]
 \center{\epsfig{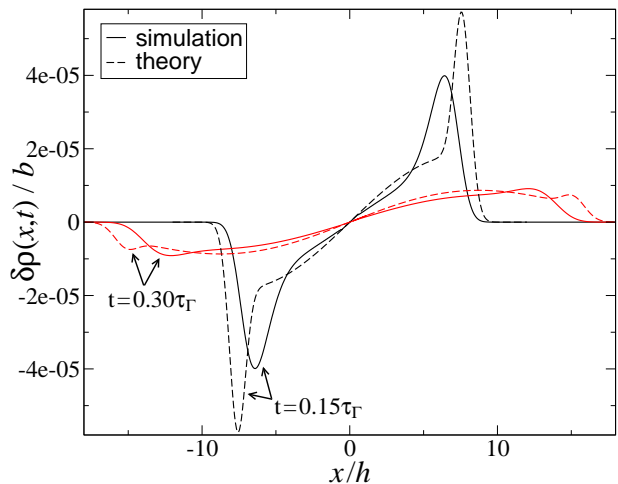}}
 \caption{Two snapshots of a density perturbation for a fluid in a square 
   channel with no-slip boundary conditions. An impulsive force is
   applied along the channel axis in the positive $x$ direction.
   Lattice--Boltzmann simulation results (where the force is applied
   to a rigid sphere of diameter $\sigma$) are compared with the
   analytical ones for a force distributed uniformly over the square
   cross-section of side $h=1.5\sigma$ (Fourier-inverted
   Eq.~(\ref{eq:rhoxi_k}), multiplied by a factor of $2$
   \cite{factor2}).  The time unit is $\tau_{\Gamma}=h^2/\Gamma$.}
 \label{fig:rhot_Q1D_impulse_noslip}
\end{figure}

Figure~\ref{fig:rhot_Q1D_impulse_noslip} shows snapshots of the
density perturbation propagating in a square channel with no-slip
boundary conditions.  The wave-like propagation quickly disappears and
the perturbation continues to evolve diffusively. Once again, the
simplified 1D theory is found to qualitatively reproduce all the
features seen in the lattice--Boltzmann simulation \cite{factor2}.
%At long times the functional form in Eq.~(\ref{eq:rho1D_xi}) 
%undermines the simulation results
%by a factor $\sim 2$.    

\section{Velocity cross-correlation function}
%============================================
\label{sec_correlation}

Section \ref{sec_sound} has been concerned with the dynamics of
density perturbations in a viscous fluid. We now proceed to examine
the effect of this dynamic response on particles suspended in the
fluid.

Hydrodynamics represents the fluid as a continuous medium. Its
discrete structure is revealed when a large particle immersed in the
fluid undergoes Brownian motion due to large number of collisions with
the much smaller fluid particles.  Fluid particles impart momentum to
the large particle, and the particle, in its turn, returns momentum to
the fluid by setting off large-scale fluid flows. These flows lead to
fluid-mediated correlations between the large particles.  The flows
can be decomposed, according to Eqs.~(\ref{eq:uT}) and (\ref{eq:L}),
into transverse and longitudinal components. The present work is
concerned with correlations caused by density perturbations, and so we
focus on the role of longitudinal flows. As noted in
Sec.~\ref{sec_intro}, transverse flows in a rigid channel are screened
at distances larger than the channel width and do not contribute to
long-range correlations.

\begin{figure}[tbh]
 \center{\epsfig{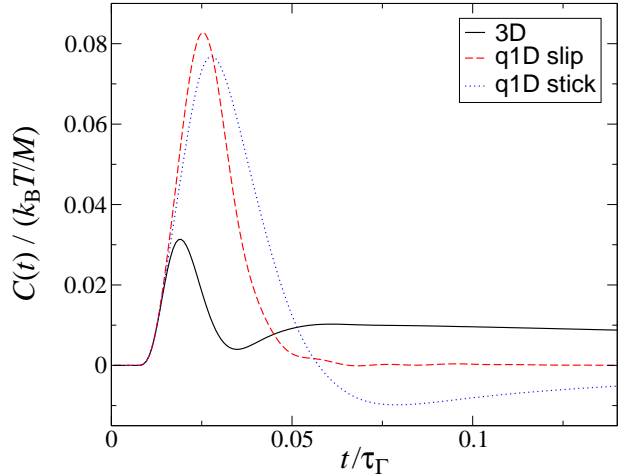}}
 \caption{Velocity cross-correlation functions  
   for two particles, as obtained from lattice--Boltzmann simulations
   of three systems: an unbounded fluid, a fluid in a square channel
   (side $h$) with slip boundary conditions, and a fluid in a square
   channel with no-slip boundary conditions. The particles have
   diameter $\sigma$ and mass $M$ and they are separated by a distance
   $d=2\sigma=1.33h$. The time unit is $\tau_{\Gamma}=h^2/\Gamma$.}
 \label{fig:v12_20}
\end{figure}

To visualize these flow-mediated correlations, we show in
Fig.~\ref{fig:v12_20} velocity cross-correlation functions between two
isolated particles, as obtained from lattice--Boltzmann simulations
for three different systems: an unbounded fluid, a fluid in a square
channel with slip boundary conditions, and a fluid in a square channel
with no-slip boundary conditions.  The cross-correlation function,
$$
C(t,d)=\Big\langle {V}_1(0){V}_2(t)\Big\rangle_{d},
$$
measures ensemble-averaged correlations between the velocities of
two particles, $V_1$ and $V_2$, separated by a distance $d$. In all
three cases we take the two velocities to be oriented along the line
that connects the centers of the two particles.  To avoid the
time-consuming fluctuating simulations and averaging procedure, we
employ instead a deterministic measurement of $C(t,d)$ based on the
linear response of the two-particle system. The equivalence of the two
measurements is guaranteed by the fluctuation-dissipation theorem
\cite{Hansen}. The procedure goes as follows. For $t<0$ both particles
are at rest, $V_1(t)=V_2(t)=0$.  At $t=0$ we apply an instantaneous
force to particle $1$ by assigning it a finite initial velocity,
$V_1(t=0)=V_0$.  As time evolves, $V_2(t)$ responds to the
perturbation set off by particle $1$. The velocity cross-correlation
function is then obtained as
$$
\frac{C(t,d)}{k_BT}=\frac{1}{M}\frac{V_2(t)}{V_1(0)},
$$
where $M$ is the particle mass and $k_BT$ the thermal energy.

\begin{figure}[tbh]
 \center{\epsfig{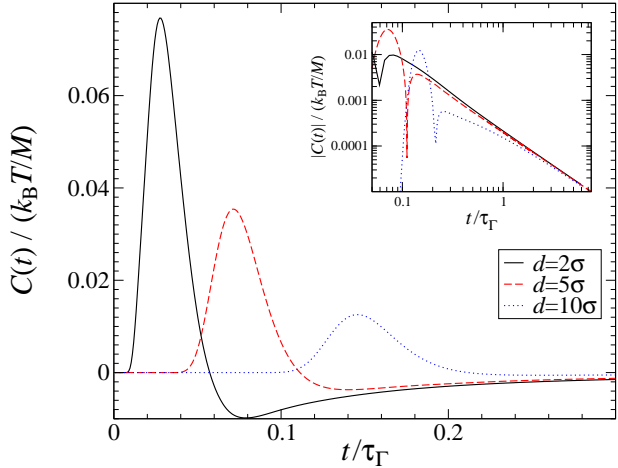}}
 \caption{Velocity cross-correlation functions for two particles in a 
   rigid channel (no-slip boundary conditions), as obtained from
   lattice--Boltzmann simulations. The particles, situated on the
   channel axis, have diameter $\sigma$ and mass $M$. The channel has
   a square cross-section of side $h=1.5\sigma$. Correlations are
   shown for three different inter-particle separations.
   %$d=1.2\sigma,2\sigma,10\sigma$.  
   The time unit is $\tau_{\Gamma}=h^2/\Gamma$. The inset shows the
   absolute value of the correlations on a logarithmic scale,
   demonstrating the $(-3/2)$ power law at long times, with a
   distance-independent amplitude.}
 \label{fig:v12_H15}
\end{figure}

In Fig.~\ref{fig:v12_20} we see that in all three systems there is a
finite incipient time required for the signal to reach particle $2$.
This time is roughly $d/c_s$. The leading signal in the channels is
larger than that in the unbounded fluid, because the perturbation is
guided along the channel rather than spreading in 3D. At larger
inter-particle distances, however, that signal is suppressed in
the case of a rigid channel with no-slip boundary conditions because
of the diffusive nature of sound. (See Fig.~\ref{fig:v12_H15}.)

The correlations at long times are very different for the three
systems. In the unbounded case the correlation is governed by a
positive algebraic tail, $\sim t^{-3/2}$. This is a manifestation of
the well known long-time tails arising from the 3D diffusion of
vorticity (monopolar transverse flow) \cite{Hansen}. In the rigid
(no-slip) channel correlations due to transverse flows are screened,
and the long-time correlation is governed by a negative algebraic
tail, $\sim(-t^{-3/2})$. This is a result of the 1D diffusion of sound
(dipolar longitudinal flow) presented in Sec.~\ref{sec_sound}. By
contrast, in the case of a channel with slip boundary conditions, the
longitudinal signal propagates acoustically, making the correlation
vanish exponentially fast once the signal has passed particle 2.

We next focus on the sound-mediated correlations in a rigid channel
with no-slip boundary conditions.  Figure~\ref{fig:v12_H15} shows
$C(t,d)$ for different separations $d$. The structure is similar for
the different separations.  There is the incipient time required for
the leading signal to reach the other particle. This signal is
suppressed with increased separation because of sound diffusion. It is
followed by a long-time negative algebraic tail, whose magnitude does
not show any dependence on separation. On the one hand, this indicates
a correlation of {\em unrestricted spatial range}. On the other hand,
the backflow associated with the negative tail causes the particle to
return almost exactly to its initial position; the time-integrated
correlation (which is equal to the coupling diffusion coefficient of
the particle pair) decays exponentially with $d/h$.  This is how the
two phenomena\,---\,long-range temporal correlations and short-range
steady-state correlations\,---\,are reconciled.

To quantitatively account for the long-time algebraic behavior of the
correlations we need to relate the diffusive density perturbation,
given by Eq.~(\ref{eq:rho1D_xi}), to the flow velocity. This can be
done either via the continuity equation, $\partial_t \delta\rho =
-\rho_0 \partial_x u$, or by applying Fick's law to the sound
diffusion, $\rho_0 u = -D_s\partial_x\delta\rho$. Both methods yield
the same long-time flow velocity,
$$
u(x,t) = -\frac{b}{4c_sh\rho_0\sqrt{\pi\alpha\nu}}
\bigg(\frac{1}{t}\bigg)^{3/2}.
$$
Equating it with the velocity advecting particle 2 and applying the
fluctuation-dissipation theorem, as explained above, we obtain the
long-time velocity cross-correlation function \cite{factor4},
\begin{equation}
C(t,d) = -\frac{k_BT}{4 c_sh\rho_0\sqrt{\pi\alpha\nu}} \bigg(\frac{1}{t}\bigg)^{3/2},
\label{eq:Clong}
\end{equation}
which has no dependence on the inter-particle separation $d$. The
negative sign of the long-time tail is traced back to the dipolar
shape of the diffusive density perturbation, Eq.~(\ref{eq:rho1D_xi}).

%For comparison, in Fig.~(\ref{fig:v12_H15_anal}) we plot velocity 
%cross--correlation functions obtained from the 1D model for the same 
%parameters,    
%%(by transforming $\delta\rho_k$ in Eq.~(\ref{eq:rhoxi_k} to $u_k$ 
%%using the conservation of mass equation and then Fourier inverting 
%%the result).  
%and in Fig.~(\ref{fig:v12_H15_water}) we use parameters corresponding
%to water.  
%\begin{figure}
% \epsfig{file=v12_H15_anal.eps, width=\columnwidth}
% \caption{Velocity cross--correlation functions between an isolated pair of 
%colloids in the channel center for three different separations: 
%$d=1.2\sigma,2\sigma,10\sigma$, obtained from the 1D model. }
% \label{fig:v12_H15_anal}
%\end{figure}
%\begin{figure}
% \epsfig{file=v12_H15_water.eps, width=\columnwidth}
% \caption{Velocity cross--correlation functions between an isolated pair of 
%colloids in the channel center and for separation $d=10\sigma$ for a liquid water, 
%obtained from the liquid water. }
% \label{fig:v12_H15_water}
%\end{figure}

\section{Collective dynamics}
%============================
\label{sec_collective}

Finally, we study the effect of sound on the collective dynamics of
many particles. The discussion in this section is restricted to the
q1D geometry of a rigid channel with no-slip boundary conditions.

We consider the wavenumber-dependent particle current correlation
function \cite{Hansen},
%\begin{equation}
%J(k,t) = \frac{1}{N}\sum_{i,j}^N
%\Bigg\langle
%\Big[{\bf\hat k}\cdot{\bf V}_i(0){\bf V}_j(t)\cdot{\bf\hat k}\Big]
%e^{i{\bf k}\cdot({\bf r}_i(0)-{\bf r}_j(t))}
%\Bigg\rangle,
%\label{eq:Jk}
%\end{equation}
%where ${\bf\hat k}={\bf k}/k$, and ${\bf V}(0){\bf V}(t)$ is the square matrix 
%with elements $V_{\alpha}(0)V_{\beta}(t)$ where $\alpha=x,y,z$, 
\begin{equation}
J(k,t) = \frac{1}{N}\sum_{i,j}^N
\Bigg\langle V_i(0)V_j(t)e^{ik(x_i(0)-x_j(t))}\Bigg\rangle,
\label{eq:Jk_1D}
\end{equation}
where $N$ is the number of particles and $i,j$ are particle labels.
Self-contributions to the sum, with $i=j$, are related to the velocity
autocorrelation function of single particles, $\langle
V(0)V(t)\rangle$. As our interest here lies solely in
cross-correlations, we subtract this contribution and consider
$J_c(k,t)=J(k,t)-\langle V(0)V(t)\rangle$.

To obtain $J_c$ from lattice--Boltzmann simulations we assume
time-scale separation between the relaxation of the fluid and that of
particle configurations. We generate configurations $X^N$ of $N$
particles using a Monte Carlo scheme. For each configuration we assign
to each particle an initial velocity from a Maxwell--Boltzmann
distribution and then follow the time evolution of every particle in a
lattice--Boltzmann simulation. This allows us to calculate, for each
configuration, the velocity cross-correlation functions between all
pairs of particles, $\langle V_i(0)V_j(t)\rangle\{X^N\}$, as described
in Sec.~\ref{sec_correlation}.  
The assumption of time-scale separation simplifies the expression for
$J_c$ to
$$
J_c(k,t) = \frac{\phi}{\sigma}\int_{-\infty}^{\infty} 
dx\, g(x)e^{ikx}\Big\langle V_1(0)V_2(t)\Big\rangle_{\rm fast}\{x\},
$$
where $g(x)$ is the equilibrium pair correlation function of the
particles, and $\phi=N\sigma/L$ ($L$ being the channel length) is their
linear fraction.  The angular brackets $\langle\dots\rangle_{\rm
  fast}$ denote a double averaging procedure: first, for a given
configuration, an average is taken over time that is sufficiently long
for the fluid to relax but sufficiently short for the configuration to
be considered ``frozen''; then another averaging is performed over
many particle configurations with the constraint $x_1-x_2=x$.

\begin{figure}[tbh]
  \center{\epsfig{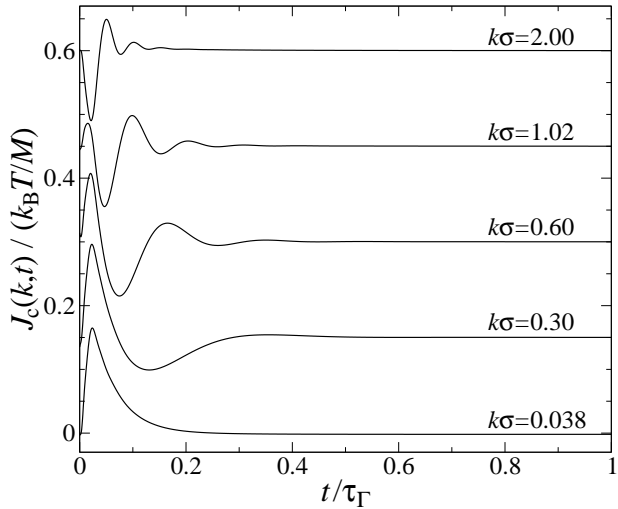}}
 \caption{Particle current correlation functions obtained from 
   lattice--Boltzmann simulations for a suspension of particles
   (diameter $\sigma$, linear fraction $\phi=0.72$) in a square
   channel (side $h=1.5\sigma$). The different curves, corresponding
   to different wavenumber $k$, are shifted vertically by $0.15$
   from each other, for clarity. The time unit is
   $\tau_\Gamma=h^2/\Gamma$.}
 \label{fig:Jc}
\end{figure}

In Fig.~\ref{fig:Jc} we plot the time evolution of $J_c(k,t)$, thus
obtained from the lattice--Boltzmann simulations, for various
wavenumbers $k$.  As $t\to 0$, $J_c$ vanishes, as the inter-particle
correlations in the compressible fluid require a finite time to
develop.  The presence of oscillations for small wavelengths,
comparable to the particle size and channel width, reflects the
wave-like propagation of sound over these small length scales. The
oscillations vanish at longer wavelenghs, indicating the onset of
diffusive sound propagation.  At long times, the particle current
correlation $J_c$ decays algebraically as $t^{-3/2}$.

To further clarify the results of Fig.~\ref{fig:Jc} we resort to a
simple model, where $g(x)$ is replaced by its dilute limit: $g(x)=0$
for $|x|<\sigma$, and $g(x)=1$ for $|x|>\sigma$.  For the velocity
cross-correlation function $\langle V_1(0)V_2(t)\rangle_{\rm
  fast}\{x\}$ we use the 1D description of Sec.~\ref{sec_correlation}.
From Eq.~(\ref{eq:rhoxi_k}) and the continuity equation,
$\partial_t\delta\rho_k=-ik\rho_0 u_k$, we obtain the flow velocity
$u_k$. The fluctuation-dissipation theorem is then used to relate the
flow to the velocity cross-correlations. This model yields
\begin{eqnarray}
\frac{J_c(k,t)}{k_BT} &=& 
\frac{\phi}{b\sigma}\left[u_k(t)-\int_{-\sigma}^{\sigma}dx\,e^{-ikx}u(x,t)\right].
\label{eq:Jcapp}
%\nonumber\\
%&+&\frac{2\phi}{b\sigma\pi}\int_{k}^{\infty} 
%dk'\,u_{k'}(t)\frac{\sin(\sigma(k'-k))}{k'-k}.
\end{eqnarray}
The qualitative change in $J_c$ as its oscillations vanish at long
wavelengths is reproduced by the first term. The onset of the
overdamped diffusive regime for small wavenumbers, lying outside the
range delineated in Eq.~(\ref{eq:limits}), should occur for the
parameters used in the simulation at $k\sigma<0.18$, which is in line
with Fig.~\ref{fig:Jc}.  The second term in Eq.~(\ref{eq:Jcapp})
supplies the function $J_c$ with the algebraic decay $\sim t^{-3/2}$,
which at long wavelength is positive.
% but changes to negative when $k$ becomes large.

%$$\lim_{k\to 0}J_c(k,t) = -\frac{k_BT\phi}{\alpha\sigma\rho_0\nu} D_sk^2e^{-D_sk^2 t},$$

Another function that characterizes the collective dynamics of
particles is the dynamic structure factor \cite{Hansen,Pusey},
\begin{equation}
S(k,t) = \frac{1}{N}\sum_{i,j}^{N}
\Big\langle e^{i k (x_i(0)-x_j(t))} \Big\rangle.
\label{eq:Sk}
\end{equation}
At times much shorter than the configurational relaxation time it
decays exponentially with time and can be written as
$$
S(k,t) = S_0(k)e^{-tk^2H(k)/S_0(k)},
$$
where $S_0(k)=S(k,t=0)$ is the static structure factor. The
hydrodynamic factor $H(k)$ accounts for the effect of fluid-mediated
correlations on the dynamic structure and is given by \cite{Pusey}
%$$
%H(k)=\frac{1}{N}\sum_{i,j}^N
%\Bigg\langle {\bf\hat k}\cdot{\bf D}_{ij}\{{\bf R}^N\}\cdot{\bf\hat k}
%\,e^{i{\bf k}\cdot({\bf r}_i-{\bf r}_j)}\Bigg\rangle_{\rm},  
%$$
%which for the 1D system simplifies to
$$
H(k)=D_{\rm self}+\frac{1}{N}\sum_{i\ne j}^N
\Bigg\langle {D}_{ij}\{X^N\}\,e^{ik(x_i-x_j)}\Bigg\rangle_{\rm},  
$$
where $D_{ij}\{X^N\}$ are the configuration-dependent pair
diffusion coefficients, and $D_{\rm self}=\langle D_{ii}\rangle$ is
the self-diffusion coefficient. Note that $H(k)$ depends only on
steady-state properties of the fluid and, hence, is unaffected by
sound.

\begin{figure}[tbh]
 \center{\epsfig{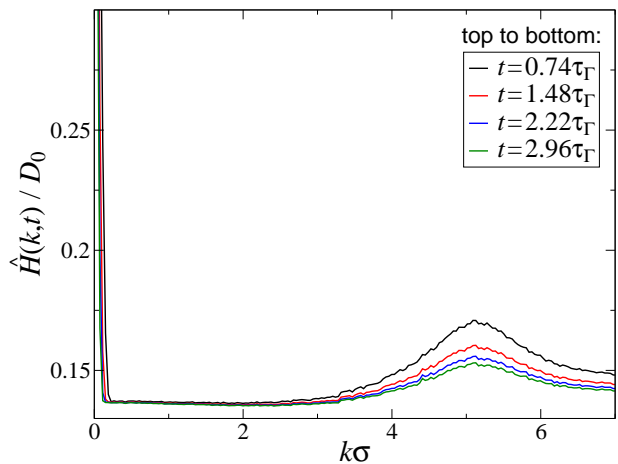}}
 \caption{Temporal hydrodynamic factor as a function of wavenumber at 
   various times. The system is the same as in Fig.~\ref{fig:Jc}.}
 \label{fig:Hqt}
\end{figure}

Recalling the Green--Kubo relation \cite{Hansen},
$$
{D}_{ij}(x)
=\int_{0}^{\infty}dt\,
\Big\langle {V}_i(0){V}_j(t)\Big\rangle\{x\},
$$
and assuming time-scale separation, we relate $H(k)$ to the
current correlation function,
$$
H(k) = \int_0^\infty dt J(k,t).
$$
To examine the effect of sound we therefore use the functions
$$
\hat H(k,t) = \int_0^{t} dt'\, J(k,t'),\ \ \ 
\hat H_c(k,t) = \int_0^{t} dt'\, J_c(k,t'),
$$
such that in the limit $t\to\infty$ $\hat H(k,t)$ reduces to
$H(k)$, and $\hat H_c(k,t)$ to $H(k)-D_{\rm self}$ \cite{ft_Ladd}.
Figure~\ref{fig:Hqt} shows the dependence of $\hat H(k,t)$ on $k$ at
various times.  As $t$ increases, $\hat H(k,t)$ converges to its
limiting form, $H(k)$, and eventually all sound effects vanish.
However, at any finite time there appears a feature that is absent in
unbounded suspensions\,---\,a sharp peak at $k\to 0$.

To see that the small $k$ feature is due to long-range sound-mediated
correlations, we return to the simplified 1D description. Using
Eq.~(\ref{eq:Jcapp}) and the continuity equation,
$\partial_t\delta\rho_k=-ik\rho_0 u_k$, we obtain
\begin{equation}
\frac{\hat H_c(k,t)}{k_BT} = \frac{\phi}{b\sigma\rho_0} \left[
\frac{\delta\rho_k(t)}{(-ik)} - \frac{1}{\pi} \int_{-\infty}^{\infty}\!\!\!dk'\,
\frac{\sin(\sigma(k'-k))}{k'-k} \frac{\delta\rho_{k'}(t)}{(-ik')} \right],
\label{eq:Hc_hat}
\end{equation}
where $\delta\rho_k$ is given by Eq.~(\ref{eq:rhoxi_k}) [or by
Eq.~(\ref{eq:rhoxi_smallk}) for small $k$]. At $t\to\infty$ this
expression vanishes, indicating that within the simplified 1D description
all steady correlations between particles confined in the channel
disappear, and $H(k)=D_{\rm self}$.  However, at any finite $t$, we
have a range of small wavenumbers, $k^2<(D_s t)^{-1}$, for which
Eq.~(\ref{eq:Hc_hat}) reduces to
\begin{equation}
\frac{\hat H_c(k,t)}{D_0} \approx \frac{3\pi\phi}{\alpha} \left(
e^{-D_s k^2 t} - \frac{\sigma}{\sqrt{\pi D_s t}} \right),
\label{eq:Hc_hat_asymp}
\end{equation}
where $D_0=k_BT/(3\pi\eta\sigma)$ is the Stokes-Einstein
self-diffusion coefficient in an unbounded fluid. Equation
(\ref{eq:Hc_hat_asymp}) contains two interesting terms originating
from sound diffusion. The first contributes at $k\to 0$ a {\em
  time-independent} constant. The second adds a {\em $k$-independent}
negative long-time algebraic tail, $\sim(-t^{-1/2})$, which derives
from the time-integrated $t^{-3/2}$ tail of the current correlation
function $J_c(k,t)$.

%$$
%\frac{\hat H(k,t)}{k_BT} = \frac{\phi}{\rho_0\sigma\alpha\nu}e^{-k^2 D_s t}
%$$
%$$
%\frac{\hat H(k,t)}{D_0} = \frac{3\pi\phi}{\alpha}e^{-k^2 D_s t}
%$$
%and the wave dependent relaxation time in the limit $k\to 0$ is
%$$
%\tau(k)=\frac{1}{D_s k^2}.
%$$

\begin{figure}[tbh]
 \center{\epsfig{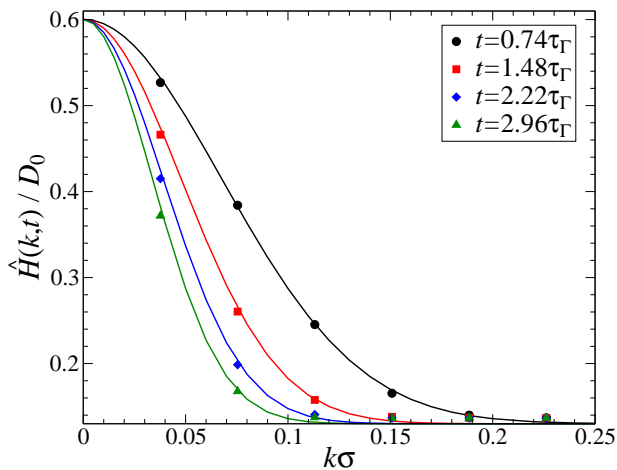}}
 \caption{Temporal hydrodynamic factor for small wavenumber 
   at various times. The system is the same as in Fig.~\ref{fig:Jc}.
   Data points represent lattice--Boltzmann simulation results. Solid
   lines show the expression $\frac{\hat H(k,t)}{D_0} =
   \frac{6\pi\phi}{\alpha}e^{-D_s k^2 t}$.}
 \label{fig:Hqt2}
\end{figure}

In Fig.~\ref{fig:Hqt2} we replot the function $\hat H(k,t)$, focusing
on the small $k$ region, and compare it with the fit $\frac{\hat
  H(k,t)}{D_0} = \frac{6\pi\phi}{\alpha}e^{-D_s k^2 t}$
\cite{factor2}. The agreement between theory and simulation is
satisfactory, both exhibiting the small-$k$ peak.

\section{Conclusion}
%===================

In the present work we have tried to establish the role of fluid
density perturbations (sound) in mediating dynamic correlations among
particles in a channel at the pair and the collective levels.  We have
been motivated in this investigation by two considerations: (i) the
hydrodynamic effects that are not associated with sound (due to
momentum and steady pressure) are screened in a rigid channel and have
no bearing on long-range particle correlations; (ii) the density
perturbations in a channel do not decay within a well defined
relaxation time (as in an unbounded fluid) but rather relax
algebraically, thus giving sound a long memory. 

The effects of sound on particle correlations in a rigid channel have
been found to be quite striking. At the two-particle level, the
velocity cross-correlation function is independent of inter-particle
distance (\ie unrestricted in its range) and has a long-time negative
algebraic tail $\sim(-t^{-3/2})$ (similar to the single-particle
autocorrelation function \cite{Frenkel97,Frenkel98,Frenkel99}). At the
collective level of a large particle assembly (q1D suspension), the
particle current correlation function has a similar long-time
algebraic tail.  We have characterized the collective correlations by
a time-dependent hydrodynamic factor, $\hat H_c(k,t)$, which converges
to the ordinary hydrodynamic factor at infinite time \cite{ft_Ladd}.
The sound-mediated correlations in the channel make this function
exhibit an anomalous peak at small wavenumbers and a long-time
algebraic temporal decay $\sim(-t^{-1/2})$. Thus, in summary, the
common picture of negligible, exponentially screened hydrodynamic
interactions among particles in a rigid channel is somewhat
misleading; strictly speaking, it applies only at steady state.

At the same time, the predicted sound-mediated correlations are weak
and may be measurable only over sufficiently short times. The
correlation is inversely proportional to the speed of sound [see, \eg
Eq.~(\ref{eq:Clong})], which makes its amplitude small, and the sound
diffusion is fast\,---\,for water in a micron-wide channel, $D_s=c_s^2
h^2/(\alpha\nu)\sim 0.1$ m$^2$/s. For example, let us examine the
correlation between the observable Brownian displacements of two
distant particles along the channel, as characterized by their pair
diffusivity,
\begin{equation}
  D_{12}(t) \equiv \frac{\langle\Delta x_1\Delta x_2\rangle}{2t}
  = \frac{1}{t} \int_0^t dt' \int_0^{t'} dt'' C(t'',d)
  = \frac{k_BT}{c_s h \rho_0 \sqrt{\pi\alpha\nu}} t^{-1/2}.
\label{D12}
\end{equation}
For a micron-wide channel filled with water, according to Eq.\ 
(\ref{D12}), measuring $D_{12}$ of order $10^{-3}$ $\mu$m$^2$/s would
require a temporal resolution of order $10^{-7}$ s. Thus, the
predicted sound-mediated effects might be observable using
experimental techniques of high temporal and spatial resolution, such
as diffusive wave spectroscopy \cite{Ladd1995} or fast optical
tracking \cite{Henderson2002,Huang2011}. The fact that $D_{12}$ is
independent of inter-particle distance may greatly facilitate the
acquisition of statistics in such fast-tracking experiments.

The short-time relevance of sound-mediated correlations was noted in
earlier works on unconfined suspensions
\cite{Ladd1995,Bakker2002,Henderson2002}. We end by underlining the
key differences between those correlations and the ones in a rigid
channel addressed here. First, whereas in an unconfined suspension
sound propagates as an underdamped wave until it is scattered by a
sufficient number of particles \cite{Ladd1995}, in a channel, because
of scattering from the boundaries, sound becomes diffusive as soon as
the distance gets much larger than the channel width. Second, in an
unconfined suspension sound-mediated correlations are a sub-dominant
effect, quickly taken over by vorticity diffusion (transverse modes)
\cite{Ladd1995,Bakker2002}. By contrast, in a rigid channel
correlations due to transverse flow are cut off, leaving longitudinal
flow as the sole mechanism of long-range correlations. Thus, any
correlation which could be resolved at inter-particle distances much
larger than the channel width should be associated with sound-mediated
effects.

% Acknowledgments
\ack
This research has been supported by the Israel Science Foundation (Grants No.\
588/06 and No.\ 8/10).

%\appendix
%\section*{Appendix}
%\setcounter{section}{1}

\section*{References and Notes}


\begin{thebibliography}{99}
\bibitem{factor2} 
  The fields (density, flow velocity) in the 1D model correspond to
  cross-sectional averages of the actual fields, whereas in the
  simulation we measure quantities at the center of the channel. This
  makes the simulation result larger than the analytical one by a
  factor of $\approx 2$. (For a parabolic velocity profile in a
  circular tube the cross-sectional average is exactly $1/2$ the
  central value \cite{FelderhofJCP11}.)
\bibitem{factor4} 
  Because of cross-sectional averaging \cite{factor2}, this result of
  the 1D model is smaller than the corresponding long-time behavior in
  the simulations by a factor of $\approx 4$.  The additional factor
  of $\approx 2$ arises from the continuity equation (or Fick's law),
  relating the cross-sectional averages of density (a roughly uniform
  transverse profile) and velocity (a roughly parabolic profile).
\end{thebibliography}
\end{document}